# Integrated Green Cloud Computing Architecture

Mohammad Naiim Hulkury
Department of Computer Science and Engineering
University of Mauritius
Reduit, Mauritius
naiim.sultanah@gmail.com

Mohammad Razvi Doomun
Department of Computer Science and Engineering
University of Mauritius
Reduit, Mauritius
r.doomun@uom.ac.mu

*Abstract*— **Arbitrary usage of cloud computing, either private or public, can lead to uneconomical energy consumption in data processing, storage and communication. Hence, green cloud computing solutions aim not only to save energy but also reduce operational costs and carbon footprints on the environment. In this paper, an Integrated Green Cloud Architecture (IGCA) is proposed that comprises of a client-oriented Green Cloud Middleware to assist managers in better overseeing and configuring their overall access to cloud services in the greenest or most energy-efficient way. Decision making, whether to use local machine processing, private or public clouds, is smartly handled by the middleware using predefined system specifications such as service level agreement (SLA), Quality of service (QoS), equipment specifications and job description provided by IT department. Analytical model is used to show the feasibility to achieve efficient energy consumption while choosing between local, private and public Cloud service provider (CSP).**

I. INTRODUCTION

Cloud computing is a set of network based services that provides on-demand, ubiquitous, convenient network access to a shared pool of configurable computing resources that can be rapidly provisioned and released in a simple and universal way with minimal management effort or Cloud Service Provider (CSP) interaction [1][2][3][4]. It offers new service opportunities with more efficient resource utilization, on-demand scalability, and cost reduction for organizations. Cloud computing has rapidly evolved through different phases that comprise of grid and utility computing and application service provisioning [5]. These services can be scaled up and down depending on the client's variable operation's needs, ensuring maximum cost efficiency. Adoption of cloud-based services enable companies to keep pace with the fast evolving and dynamic business climate, as well as benefiting better reliability, less maintenance and higher accessibility.

Traditional access practice to cloud services is often not energy efficient [6]. Companies using cloud often lacks policies and decision making rules to enforce the overall management and reduction of energy consumption, thus relying only on ad-hoc end-user's responsiveness to use the energy-saving schemes installed in their computing infrastructure. Moreover, a cutback in the energy budget of a data centre, without sacrificing service level agreements, is an important objective as well as economic incentive for data centre operators [7]. In this work, we focus on optimizing client-side requests to cloud services, in order to achieve lower energy consumption locally, thereby protecting the environment with lower carbon footprint when access cloud services. We use a simplified energy consumption model based on cloud services accessed for a particular job. The solution proposed is an Integrated Green Cloud Architecture (IGCA). It includes a smart client-oriented 'Green Cloud Middleware' that will smooth the transition from conventional (local) computing to cloud computing using lowest energy consumption, but subject to satisfying the service level agreement, maintaining the quality of service and staying within the budget.

The rest of this paper is structured as follows: Section 2 described related works in energy-efficient approach for cloud access. The proposed integrated green cloud architecture and evaluation are presented in section 3 and 4, respectively. Finally, section 5 concludes the work.

II. RELATED WORKS

Cloud computing uses metering technique for managing and optimizing its service and provides reporting and billing information to customers. Private cloud computing is deployed within the company premises having own data centre. Company gains higher flexibility of configuration and expansion. Public clouds are normally deployed off customer premises and run, managed and maintained by CSPs. They combine applications from different customers and provide maximum performance-cost value through a homogeneous and hi-scale methodology on shared infrastructure. Hybrid cloud supplements the private cloud with the resources of a public cloud to alleviate the processing in private cloud. However, it is critical to decide how to optimally distribute data across public and private cloud.

Cloud computing services are deployed based on business models and requirements. *Platform as a Service (PaaS)* offers clients a development platform to set-up their own software and applications. The platform's network architecture and its running operating systems are solely managed by the CSP. *Software as a Service (SaaS)* offers client on-demand access to a wide range of applications and services that are hosted in the cloud. Client using this cloud service experience a considerable saving in upfront cost of procuring new software and infrastructure. While all the installations and maintenances are done at the CSP premises, the client need just to hire high bandwidth network connectivity to the cloud. *Infrastructure as a Service (IaaS)* is composed of virtual and physical servers, storage and clusters. The infrastructure is pre-configured with storage and programming environment according to the clients' needs. Client also has control to manage the systems in terms of the operating systems, applications, storage, and network connectivity.

In [8], an approach is suggested for accessing the optimal cloud service in the federated cloud computing environment by matching the capability and performance of the foreign cloud service providers with the requirements of the client. In [9] [10][ 11], several research have proposed solutions in reducing energy consumption in cloud datacenter through virtualisation, server consolidation, and intelligent cooling systems. A *GreenCloud* Architecture is proposed in [12] with inclusive online-monitoring, live virtual machine migration, and VM placement optimization. Green Open Cloud architecture [13] is designed by taking into account the energy usage of virtualization frameworks. A simulation environment for energy-aware cloud computing data centers is described in [14] which also models communication patterns of the energy consumed by data center components.

But, there is a need to investigate about the energy consumed in accessing the cloud [15][11] and most importantly: *How to select cloud access (public or private) the greenest way without performance degradation from the client-edge?* In [16], power/energy profiling framework is presented for collecting and analyzing energy consumption data in compute clouds at CSP. In [19] an evaluation is performed of a green scheduling agorithm for energy savings in cloud Computing on CSP side only. Baliga et al. [1] presented a technical analysis of energy consumption in data transmission, data processing and data storage in both public and private clouds. The energy consumed in transmission and switching in both public and private clouds, contributes significantly to the overall cloud energy consumption. While cloud computing can be very efficient for less intensive and infrequent task, it can yet in some circumstances consume more energy than conventional computers. Cloud computing adoption is not always a 'green' solution [17] and *Table I* summarizes the conditions under which energy consumption is significant in transport, storage and processing in public and private clouds [1].

It is feasible to determine the carbon emission in a private cloud by measurements and computation, whereas that of public clouds is unknown since the public cloud architecture and access medium are transparent to the client. It is therefore problematic for a client to choose the most energy efficient access to the 'greenest cloud'. To resolve this issue, the authors in [18] proposed a green cloud architecture, whereby a client first has to send its request to a middleware green broker. The latter manages the selection of the 'greenest cloud' provider to serve client's request for cloud service. The green cloud broker in [18] depends on (i) the Carbon Emission Directory (CED) that conserves all the data related to energy efficiency of cloud services and (ii) the Green Offer Directory (GOD) that contains green services, pricing and the time it could be accessed for minimum carbon emission. The access to any public or private cloud outside the user premises is therefore guaranteed to be the greenest possible way if the client makes use of the green broker. However, a unified solution is required that takes into consideration energy consumption of jobs at client-side, CSPs CED, QoS, SLA and security issues.

TABLE I. CONDITIONS UNDER WHICH ENERGY CONSUMPTION IS SIGNIFICANT [1]

| Cloud Services | Energy component | Public cloud | Private cloud |
|---|---|---|---|
| Software as a Service | *Transport* | High Frame rates | Never |
| | *Storage* | Never | Never |
| | *Processing* | Few users per server | Few users per server |
| Storage as a Service | *Transport* | Always | High download rates |
| | *Storage* | Low download rates | Low download rates |
| | *Processing* | Never | High download rates |
| Processing as a Service | *Transport* | Medium to high encodings per week | Never |
| | *Storage* | - | - |
| | *Processing* | Medium to high encodings per week | Medium to high encodings per week |

III. INTEGRATED GREEN CLOUD ARCHITECTURE

We propose an Integrated Green Cloud Architecture (IGCA), the detail illustrated in figure 1, which consists of: the client (e.g. can be a company), a client-oriented green cloud middleware and the green broker, similar to the one in [18]. The green cloud middleware provide the client a tool to better manage the distribution of tasks to cloud with the least carbon emission (i.e. least power consumption) and other relevant decision criteria. The middleware is composed of a user interface application and a windows service.

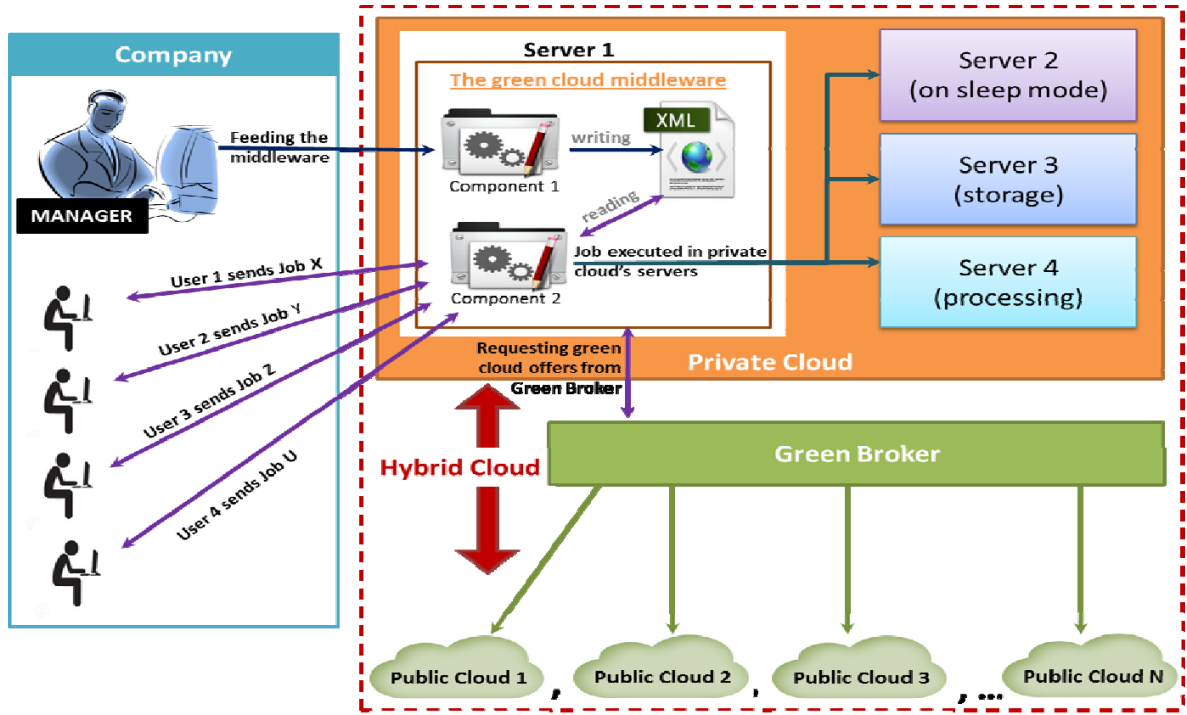

Figure 1. Integrated Green Cloud Architecture.

*A. Green Cloud Middleware: User Interface Application*

The user interface application provides a facility for the client-side (e.g. Company) to input the required specifications to the system. The specifications of the company's private cloud are vital information requirement in the IGCA. For instance, An IT manager will input these detail information in the middleware. The main function of each server (e.g. $S_2$, $S_3$ & $S_4$) in the private cloud will be specified such as for storage, processing and backup purpose. The usage frequency ($F_i$) of each server will be identified as 'rarely used', 'intermittently used' or 'continuously' accessed/used. The mode of each server can be of types: heavy duty ($Hv$), sleep mode ($Sl$) or hibernate ($Hi$). Other relevant information of the private cloud, such as storage capacity ($C$), processing power ($P$), and energy consumption ($E$) are also available and provided to the middleware.

Similarly, the client PC's specifications, such as storage and processing capacity and power consumption rating, are also registered in the middleware. Middleware component has the comprehensive technical information regarding the private cloud and specifications of the local computer machines. IT (project or operation) management level involved with business operations and decision making generally performs break down schedule of the business operations into jobs and sub jobs. Each job/ sub-job has specific processing requirement and IT resource needs. The description, name/title, associated budget, SLA, and QoS of different jobs are typical information that the manager will input in the middleware upon creation of a job. Then, the carbon emission for the job execution on a client's PC, on a server in the private cloud and on a server in the public cloud are estimated and displayed to the user. The greenest option is thus proposed but yet allowing the manager to choose where to execute a particular job based on: SLA, security level requirement, execution carbon emission and budget associated with the job.

As example, a particular job proposed to be executed in a public cloud may not be the best option as its security level may be high or its SLA's or budget does not permit it to be executed in a public cloud. In this context, the manager would take his decision based on the two remaining options (Local host or private cloud). Once the choice is finalised for a job, this information is saved in an XML file residing on the server where the middleware is installed. After configuring each job execution's destination, a table matching the job and its destination's execution address is saved in the XML file. The address to a server in the public cloud is dynamic. It is during execution that the middleware will fetch the best green path with the green broker.

The green broker, proposed by Buyya R. et al [3], as stated earlier relies on two main components, the Carbon Emission Directory (CED) and Green Cloud Offers (GCO). The green cloud broker will initially use current energy parameters in the CED and calculates the carbon emission of all the CSPs who are offering the demanded cloud service.

Based on obtained results, the green broker will buy the cloud offers with the least carbon emission on behalf of the user.

### B. Green Cloud Middleware: Windows Service

The second component of the Green Cloud Middleware is a windows service running on the *server 1* residing the XML file. The service listens to all jobs executions' requests from any client. If the job is not known to the middleware, a notification is sent to the client and to its job manager. In this situation, the job execution will be performed either on the client's PC or on a server in the company's private cloud. If the job has been registered in the middleware, the latter will read its corresponding execution address from the XML file. For LOCALHOST address, the notification is sent to the request's sender that will in turn execute the job locally. This notification intelligence is also stored at the client PC so as next time the PC can automatically execute identical job request without communicating with the Middleware. For jobs to be executed in the private cloud, the address configured with the middleware includes the server's name. Job with corresponding execution address to public cloud, the service will communicate with the green broker to decide the best green option offer for the job.

As one of the requirements of the middleware, the management team (or IT manager) will input information about the company private cloud. The information of the PC's that the employees use in each department or team in the company is essential for the middleware to calculate energy consumption of jobs at local host, as will be shown by energy model equations. The main information fed to the middleware are:

*[Job ID, Job Name, Dept, Frequency, Job Size, Bandwidth Req]* $\in$ **Job description**

*[Security Level, QoS, Budget]* $\in$ **Service Level Agreement and Contract**

*[Communication cost, Bandwidth]* $\in$ **Network Specifications**

*[Processor speed, Hard disk capacity, Memory usage, Average Power consumption idle/non-idle]* $\in$ **PC Specifications, Server Specifications**

Decision for Cloud allocation depends on several factors such as: Job description, SLA, network specifications as well as Energy Consumption per byte or Job (for Local, Private, Public), Security level policy, Cost of service, Computation or processing speed, bandwidth provisioning. Therefore, estimated energy consumption for the job workload is computed and compared for the three options. For each job, it is assumed that the amount of data to be transferred in bytes and the computational requirement in terms of CPU cycles are available.

$E\_ST_{Cloud}$, the energy consumed using cloud for a **storage service** of a particular job is modelled as:

$$E\_ST_{Cloud} = S_{file} \times N_{Download/hour} \times (E_{Transport} + E_{ContentServer}) \times U_{NumOfUsers}$$

$E\_ST_{Local}$, the energy consumed for storage or softare service on local machine is:
$$E\_ST_{Local} = E_{LocalMachine} + (S_{File} \times E_{HardDisk})$$

$E\_SF_{Cloud}$, the energy consumed in cloud for a **software service** of a particular job is given as:
$$E\_SF_{Cloud} = E_{LocalMachine} + E_{Server/user} + (F_{FrameRate} \times E_{Transport}) + (S_{File} \times E_{ServerHardDisk})$$

Energy consumed in cloud, $E\_PR_{Cloud}$, for a **processing service** of a particular job per week as a function of the number of encodings is:
$$E\_PR_{PrivateCloud} = (E_{LocalMachine} \times H_{AverageHour/week}) + (N_{NumberOfEncoding} \times H_{AverageHour/Encoding} \times E_{Server}) + (F_{FrameRate} \times E_{Transport})$$

Finally, $E\_PR_{Local}$, the energy consumed by the same job per week as a function of the number of encodings on local machine for **processing service** is:
$$E\_PR_{Local} = E_{LocalMachine} \times H_{AverageHour/week}$$

In the above equations, $S_{File}$ is average size of file, $N_{Download/hour}$ is number of download per second of the file, $E_{Transport}$ is energy consumed in transporting the file on the network, $E_{ContentServer}$ is energy consumed on content server, and $U_{NumOfUsers}$ is number of users for this job, $E_{LocalMachine}$ is energy consumed by local machine, $E_{HardDisk}$ is energy consumption of hard disk, $E_{Server/user}$ is energy consumed on server per user, $F_{FrameRate}$ is frame rate, $E_{ServerHardDisk}$ is energy consumed by server hard disk, $H_{AverageHour/week}$ is average hours per week the local machine is used, $N_{NumberOfEncoding}$ is the number of encoding per week, and $H_{AverageHour/Encoding}$ is the average number of hours it takes to perform one encoding.

The three outcomes are summarized as:

{$E_{PublicCloud}$ < ($E_{Local}$, $E_{PrivateCloud}$)} & Job {SLA, QoS, Budget} compliance

{$E_{PrivateCloud}$ < ($E_{local}$, $E_{PublicCloud}$)} & Job {SLA, QoS, Budget} compliance

{$E_{Local}$ < ($E_{PrivateCloud}$, $E_{PublicCloud}$)} & Job {SLA, QoS, Budget} compliance

## IV. IGCA APPRAISAL

The proposed IGCA when adopted can help companies go 'green' and restructure their access to cloud in a more energy efficient way. The middleware is a useful tool for companies adopting cloud computing as a solution for their IT resources. It also gives companies smarter control on their cloud service requests and the flexibility to restructure and manage their cloud access accordingly, e.g. a change in SLA for a particular job. Decision making is taken care by middleware using well-defined system specifications with the aim to accomplish least energy consumption at client and CSP.

The middleware is normally installed on a server of the company's private cloud, as illustrated in Figure 2. It has several interface screens which are accessed by users directly from their PC. There is also a service which runs on the server listening to request from clients. The client side has a software or add-ins establishing the communication with this service. The middleware can also communicate with application running on different platform. Thus, the middleware contributes to make better decision for minimizing carbon footprint in cloud architecture (client and CSP side). It is reasonable to assume that the middleware itself, with its processing, has negligible energy consumption. A lightweight middleware development is necessary by employing green software development techniques.

*4.1 Case Study*

The goal of our case study is to demonstrate how IGCA satisfies green cloud solution on client-side and meet enterprise requirements to select appropriate local or cloud access for energy-efficient job execution. AXY company has a data center that consists of 3 small switches, 1 Ethernet switch and 1 router, with specifications detailed in Table II.

TABLE II. EQUIPMENT POWER SPECIFICATION

| Equipment | Make | Capacity Gb/s | Power consumption (Watts) |
|---|---|---|---|
| Small switches | Cisco 4503 | 64 | 474 W |
| Big switches (Ethernet) | Cisco 6509 | 160 | 3800 W |
| Router | Juniper MX-960 | 660 | 5100 W |

The private cloud consists of two HP DL380 G5 servers having a capacity of 800 Mb/s, a power consumption of 225 Watts and has a hard disk of 500 Gbytes capacity and having a power consumption of 2.5 Watts. The company has a single department with PCs running 32-bit, having a power consumption of 210 Watts and a hard disk of 20 Gbytes of 0.25 Watts power consumption. The screen resolution of each PC is 1280 x 1024 with a 24-bit colour. The estimated power consumed per bit in transport over the internet is 2746.38 Watts.

Three scenarios are summarized in Table III, showing the amount of power consumption for different job types.

**Scenario 1:** *Job J-STORAGE requires a storage service for a file of 1GB. The job is performed by 5 employees and is to be downloaded twice per hour.*

Since the number of download is only twice per hour, the energy consumption at private cloud is low compare to local machine. However, as this ratio rises, the power consumption to private cloud and public cloud also rises (i.e for 20 downloads per hour, the power consumption in private cloud become 2.69 W and that of public cloud rises to 15.26 W).

**Scenario 2:** *Job J-SOFTWARE requires a software service for 10-200 employees having an average of 5 GB of data for storage.*

As the number of users increases to say 200, the power consumption in local machine remain the same while the power consumption at private cloud slightly reduce to 15229.54 W while that of public cloud is estimated to 86517.72 W. However, if the same test scenario is performed with 200 employees and a frame rate of 11.5 Mb/s, the power consumption at local machine remain the same but, that at the private cloud reduces to 5564.29 (a reduction of 36.5%) and 31590.12W on public cloud. It shows that the power consumption for software service is highly dependent to frame rate.

**Scenario 3:** *Job J-PROCESSING requires a big processing service of a video file of 10 GB average size for 1 week of 40 hours of office time with 20 encodings per week.*

Local machine always have the least power consumption while power consumption in both public and private cloud is very high due to transport cost.

TABLE III. CLOUD POWER CONSUMPTION V/S JOB TYPE

| Job Type | Infrastructure | Power consumption (Watts) |
|---|---|---|
| *Job J-Storage* | Local Machine | **6.75** |
| | Private Cloud | **0.27** |
| | Public Cloud | **1.53** |
| *Job J-Software* | Local Machine | **6.75** |
| | Private Cloud | **15229.61** |
| | Public Cloud | **86517.76** |
| *Job J-Processing* | Local Machine | **262.5** |
| | Private Cloud | **7255434.17** |
| | Public Cloud | **19221342.96** |

The different scenarios representing each jobs of company AXY are saved with the choice of execution destination made by the manager into an XML file. In order to test the mapping of job to execution destination supposed to be performed jointly by the user software add-in, the windows service of component 2 and the green broker, a modelling tool has been used simulating all the three former constituents. The choices of the manager were strictly respected by the tool and all additional modifications done on a job (i.e changes in SLAs) taken into consideration correctly. IGCA is an attractive mechanism to reduce the overall energy consumption at the "client edge", as well as proposing the most energy-efficient hybrid cloud support for executing jobs with the required QoS and SLA to ensure the customer satisfaction. The middleware takes into consideration fine-grained characteristics such as job description and hardware heterogeneity, networking and server's capacity that exist in real cloud scenario.

V. CONCLUSION

The energy efficiency of cloud computing has become one of most pressing issues. An Integrated Green Cloud Architecture with a Middleware component enables company manager to balance the job operations' execution efficiently in terms of minimum energy consumption to private or public clouds or simply based on user's request. IGCA provide a client oriented green cloud middleware allowing the job/operation manager to define the job of each department, specifying the related QoS level and the SLAs involved. Energy consumption of job execution on private cloud, public cloud and local host is pre-determined based on modelling scenario. Green cloud computing initiatives will be vital when it comes to offloading workload to clouds (public/private) while providing the best energy-performance value.

ACKNOWLEDGMENT

The authors thank anonymous reviewers for their comments to improve the paper.

REFERENCES

[1] J. Baliga, R. Ayre, K. Hinton, and R.S. Tucker, "Green Cloud computing: Balancing energy in processing, storage and transport", Proceedings of the IEEE, 99(1) 149-167, January 2011.

[2] L. Wang and G. Von Laszewski, "Scientific Cloud Computing: Early Definition and Experience", Proceedings of the 10th IEEE International Conference on High Performance Computing and Communications, 2008.

[3] Mell P. and Granc T.( September 2011), "The NIST Definition of Cloud Computing", National Institute of Standards and Technology (NIST). [Online]. Available: http://csrc.nist.gov/publications/nistpubs/800-145/SP800-145.pdf

[4] Souvik Pal and Prasant Kumar Pattnaik, "Efficient Architectural Framework for Cloud Computing", *International Journal of Cloud Computing and Services Science (IJ-CLOSER)* Vol.1, No.2, June 2012, pp. 66~73

[5] Yeo C. S., Buyya R., Dias de Assunção M., Yu J., Sulistio A., Venugopal S., and Placek M., "Utility Computing on Global Grids", Chapter 143, Hossein Bidgoli (ed.), The Handbook of Computer Networks, ISBN: 978-0-471-78461-6, John Wiley & Sons, New York, USA, 2007.

[6] Andreas Berl, Erol Gelenbe, Marco Di Girolamo, Giovanni Giuliani, Hermann De Meer,Minh Quan Dang and Kostas Pentikousis, Energy-Efficient Cloud Computing, The Computer Journal, Advance Access:1—7, 2009

[7] Rabi Prasad Padhy and Manas Ranjan Patra, "An Enterprise Cloud Model for Optimizing IT Infrastructure" International Journal of Cloud Computing and Services Science (IJ-CLOSER) Vol.1, No.3, August 2012, pp. 123~133

[8] Saumitra Baleshwar Govil, T. Karthik, S. Karthikeyan, Vijay K. Chaurasiya, Santanu Das, "An Approach to Identify the Optimal Cloud in Cloud Federation", International Journal of Cloud Computing and Services Science (IJ-CLOSER) Vol.1, No.1, March 2012, pp. 35 ~ 44

[9] Y.C. Jin, Y.G. Wen and Q.H. Chen, *"Energy Efficiency and Server Virtualization in Data Centers: An Empirical Investigation,"* Proceedings of IEEE INFOCOM Workshop on Communications and Control for Sustainable Energy Systems: Green Networking and Smart Grids (co-host with Infocom'12), March 25-30, 2012, Orlando, Florida, USA.

[10] Suzanne Niles and Patrick Donovan. "Virtualization and Cloud Computing: Optimized Power, Cooling, and Management Maximizes Benefits". White paper 118. Revision 3, Schneider Electri, 2011.

[11] Anton Beloglazov, Rajkumar Buyya, Young Choon Lee, and Albert Zomaya, "A Taxonomy and Survey of Energy-Efficient Data Centers and Cloud Computing Systems" *Advances In Computers, Vol. 82*

[12] Liang Liu, Hao Wang, Xue Liu, Xing Jin, WenBo He, QingBo Wang, Ying Chen, "GreenCloud: A New Architecture for Green Data Center" ICAC-INDST'09, June 16, 2009, Barcelona, Spain. Pp. 29-38

[13] Anne-Cécile Orgerie and Laurent Lefèvre, "When Clouds become Green: the Green Open Cloud Architecture" in proceedings International Conference on Parallel Computing (ParCo) 2009

[14] D. Kliazovich, P. Bouvry, S. Ullah Khan, "GreenCloud: a packet-level simulator of energy-aware cloud computing data centers" J Supercomput DOI 10.1007/s11227-010-0504-1

[15] Ismael Solis Moreno, Jie Xu, "Energy-Efficiency in Cloud Computing Environments: Towards Energy Savings without Performance Degradation" IJCAC 2011, Vol. 1 No. 1, p.17-33.

[16] Z. Zhang and S. Fu, "Profiling and Analysis of Power Consumption for Virtualized Systems and Applications", IPCCC 2010: 329-330.

[17] A.-H. Mohsenian-Rad and A. Leon-Garci, "Energy-Information Transmission Tradeoff in Green Cloud Computing," *Proc. IEEE GlobeCom '10,* Mar. 2010.

[18] S.K. Garg, C.S. Yeo and R. Buyya," Green Cloud Framework For Improving Carbon Eciency of Clouds", Proceedings of the 17th International European Conference on Parallel and Distributed Computing (EuroPar 2011, LNCS, Springer, Germany), Bordeaux, France, August 29-September 2, 2011.

[19] T.V.T. Duy, Y. Sato, and Y. Inoguchi, "Performance evaluation of a Green Scheduling Algorithm for energy savings in Cloud computing", ;in Proc. IPDPS Workshops, 2010, pp.1-8.